\documentclass[conference]{IEEEtran}
\IEEEoverridecommandlockouts

\usepackage{cite}
\usepackage{amsmath,amssymb,amsfonts}
\usepackage{algorithmic}
\usepackage{graphicx}
\usepackage{textcomp}
\usepackage{xcolor}
\usepackage{hyperref}
\usepackage{url}
\usepackage{tikz}

\def\BibTeX{{\rm B\kern-.05em{\sc i\kern-.025em b}\kern-.08em
    T\kern-.1667em\lower.7ex\hbox{E}\kern-.125emX}}

\newcommand\copyrighttext{%
  \footnotesize This work has been submitted to the IEEE for possible publication. Copyright may be transferred without notice, after which this version may no longer be accessible.}
\newcommand\copyrightnotice{%
\begin{tikzpicture}[remember picture,overlay]
\node[anchor=south,yshift=11pt] at (current page.south) {\fbox{\parbox{\dimexpr\textwidth-\fboxsep-\fboxrule\relax}{\copyrighttext}}};
\end{tikzpicture}%
}
    
\begin{document}

\title{From RTL to Prompt Coding: Empowering the Next Generation of Chip Designers through LLMs
\thanks{This paper was funded by the German Federal Ministry of Research, Technology and Space (BMFTR) as part of the ``Chipdesign Germany'' project under grant number 16ME0890.}
}

\author{\IEEEauthorblockN{Lukas Krupp\IEEEauthorrefmark{1}, Matthew Venn\IEEEauthorrefmark{2} and Norbert Wehn\IEEEauthorrefmark{1}}
\IEEEauthorblockA{\IEEEauthorrefmark{1}RPTU University of Kaiserslautern-Landau, Kaiserslautern, Germany\\
\IEEEauthorrefmark{2}Tiny Tapeout}
}

\maketitle
\IEEEpubid{\begin{minipage}{\textwidth}
  \copyrightnotice
\end{minipage}} 

\begin{abstract}
This paper presents an LLM-based learning platform for chip design education, aiming to make chip design accessible to beginners without overwhelming them with technical complexity. It represents the first educational platform that assists learners holistically across both frontend and backend design. The proposed approach integrates an LLM-based chat agent into a browser-based workflow built upon the Tiny Tapeout ecosystem. The workflow guides users from an initial design idea through RTL code generation to a tapeout-ready chip. To evaluate the concept, a case study was conducted with 18 high-school students. Within a 90-minute session they developed eight functional VGA chip designs in a 130 nm technology. Despite having no prior experience in chip design, all groups successfully implemented tapeout-ready projects. The results demonstrate the feasibility and educational impact of LLM-assisted chip design, highlighting its potential to attract and inspire early learners and significantly broaden the target audience for the field.
\end{abstract}


\section{Introduction}
The shortage of skilled chip designers has emerged as a pressing issue, particularly in Europe but also in countries like the United States and China. Recent reports highlight a growing gap between industry demand and the available workforce \cite{b1, b2}. An alarming trend is the declining interest among students in microelectronics, further worsening the designer shortage if no actions are taken \cite{b3}. On the other hand, enthusiasm for software engineering and artificial intelligence (AI) has been steadily increasing over the last years. 

A main reason for this divergence lies in the differing productivity dynamics of the two domains. Software development and AI benefit from high levels of abstraction and minimal infrastructure requirements. In contrast, chip design remains an efficiency-driven discipline where parameters such as area, power consumption and timing are critical. The resulting need for optimization across all design layers hampers high-level abstraction. Unlike software, chip design is bound by physical laws and the complexities of the chip implementation. 

Consequently, productivity improvements in hardware have largely lagged behind those in software. For instance, digital design still relies mostly on Register Transfer Level (RTL) coding using hardware description languages (HDLs) like Verilog and VHDL. HDLs allow for fine-grained control over critical physical implementation parameters but make the RTL design process time-consuming and error-prone. Furthermore, the digital design flow, from system-level specification to physical layout, requires a suite of specialized tools as well as deep expertise in design methodologies, circuits, and technology. 

Whereas software development provides a “fast path to success,” attracting learners with its immediacy, chip design confronts newcomers with a steep learning curve that can be discouraging. The design complexity, the need for specialized infrastructure, and the corresponding “long path to success” make it challenging to attract learners and maintain motivation. To increase the number of future chip designers, it is essential to spark curiosity as early as possible, ideally already at the school level \cite{b4}. To address the broad pool of non-experts, chip design must be both accessible and engaging.

The Tiny Tapeout initiative \cite{b5} has demonstrated the potential of lowering entry barriers by enabling chip design with minimal infrastructure requirements. Only a computer and internet access is needed to get started. The key lies in the use of open-source electronic design automation (EDA) tools and process development kits (PDKs), which provide unrestricted access and allow cloud-based deployment of the backend flow. Combined with low-cost manufacturing, these advances address the challenges of chip implementation and fabrication. However, frontend design knowledge, particularly in RTL coding for digital design, remains necessary and continues to pose a barrier for beginners. To date, no educational platform exists that assists learners holistically across frontend and backend design, guiding them from idea to tapeout-ready chip.

Therefore, the goal of this paper is to explore how large language models (LLMs) can be leveraged in chip design education to lower the entry barrier to RTL design for non-experts, while still enabling them to implement realistic, tapeout-ready designs. We present the following key contributions:
\begin{enumerate}
    \item An LLM-based chat agent designed to assist learners with no or only little prior knowledge in RTL coding.
    \item A methodology and browser-based environment for developing video graphics array (VGA) chip designs in a 130\,nm technology, from idea to GDSII, using the chat agent and the open-source Tiny Tapeout infrastructure.
    \item A case study with 18 high-school students that validates the methodology and demonstrates its feasibility.
\end{enumerate}

The case study designs, access to the chat agent and practical workflow instructions are provided via GitHub\footnote{Designs and workflow demo: \url{https://github.com/tukl-msd/prompt2silicon}}. 

\begin{figure*}[t!]
  \centering
  \includegraphics[width=\linewidth]{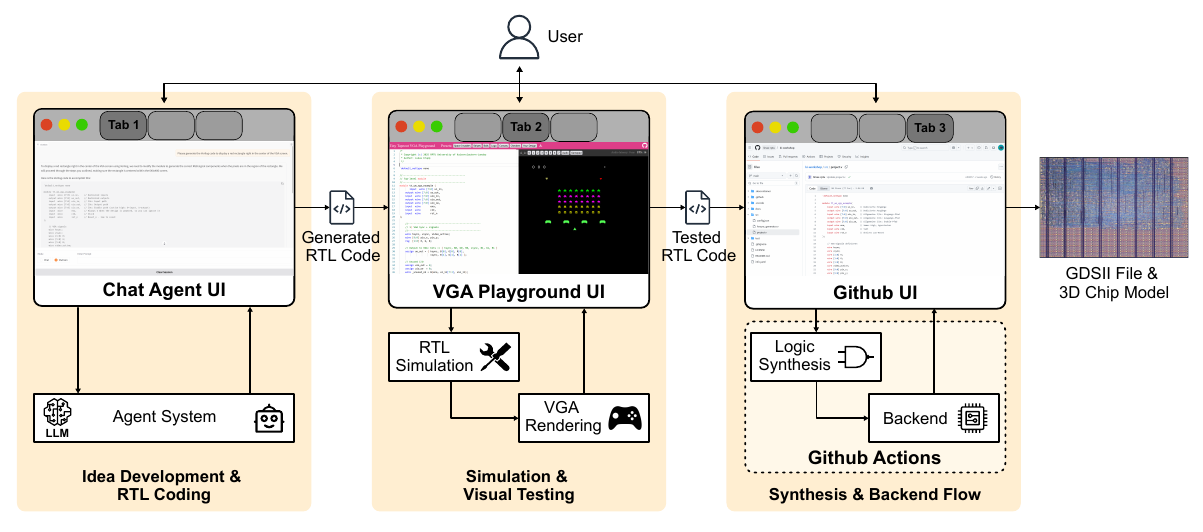}
  \caption{Overview of the proposed idea-to-GDSII learning workflow integrating the LLM-based chat agent for the RTL implementation, VGA simulation tool, and GitHub-driven backend flow. The browser-based user interfaces (UIs) and the underlying applications are shown.}
  \label{fig:workflow}
\end{figure*}

\section{Background and Related Work}
In chip design education, we distinguish three target groups. The first are advanced learners with practical experience such as PhD or Master's students, representing the smallest group. The second group comprises early or interested learners, who are intrinsically motivated to explore chip design basics. The largest group is the non-expert audience, including pupils, vocational and early Bachelor students or hobbyists, who may have heard of chip design but lack any previous exposure.

The Tiny Tapeout initiative has considerably lowered the entry barrier, particularly for interested learners and non-experts, to engage in custom chip design. Tiny Tapeout extends the established concept of multi-project wafers to multi-project dies, enabling hundreds of designs to share a single chip \cite{b6}. A multiplexer fabric allows users to select one design at a time, which is then connected to the input and output pins. Since its launch in 2022, Tiny Tapeout has facilitated the fabrication of more than 2,000 designs using open-source PDKs from SkyWater \cite{b7} and IHP \cite{b8} in 130\,nm technology nodes. 

Furthermore, Tiny Tapeout offers an extensive open-source ecosystem. Central to the ecosystem are GitHub repository templates for both digital \cite{b9} and analog designs\cite{b10}. The templates include automated synthesis and backend flows orchestrated through GitHub Actions and based entirely on open-source EDA tools. For beginners, graphical tools such as Wokwi \cite{b11} and SiliWiz \cite{b12} provide design and learning options. Other resources include visualization tools like the VGA Playground \cite{b13} and GDS Viewer \cite{b14}, as well as tutorials and videos. After tape-out, users receive their fabricated chip on an evaluation board. The system software enables out-of-the-box bring-up of the board and interaction with the designs. 

Despite these simplifications, digital design still requires substantial knowledge of RTL coding, which poses a challenge for beginners. Recent research has demonstrated the potential of LLMs to support EDA workflows \cite{b15}, showing promising results in RTL code generation, debugging, optimization, verification, and comprehension \cite{b16, b17, b18, b19, b20, b21}. However, existing efforts have primarily focused on industry-oriented use cases, where LLMs enhance the productivity of experienced designers. In contrast, the educational use of LLMs has already seen broad adoption in coding \cite{b22, b23} and STEM disciplines \cite{b24, b25}, where they serve as tutors and learning accelerators. 

Yet, to date, no LLM-based educational platforms have been introduced in the context of chip design covering the complete flow from the idea to the tapeout-ready chip.

\section{Methodology}
\subsection{System Overview \& Learning Workflow}
The human-in-the-loop design flow, shown in Fig.~\ref{fig:workflow}, enables users to interact with the system through a web browser and progress from an initial idea to a tapeout-ready chip. The workflow begins with the idea development, where users define the concept they wish to implement in the context of VGA-based designs. Inspiration can be drawn, for instance, from retro games such as \textit{Space Invaders} or \textit{Tetris}, which remain highly popular, especially among younger audiences. Designs can range from animations to interactive control elements or games by using the input pins. During this phase, prompt engineering plays a central role. Users must describe the intended behavior with sufficient precision to ensure that the generated RTL code aligns with their expectations.

Next, users interact with the LLM-based agent system through the chat agent user interface (UI), which serves as the entry point for transforming design ideas into hardware descriptions. Building on the concepts and behavioral specifications formulated during idea development, users provide their prompts to generate the RTL code. A detailed description of the underlying agent system is provided in Section~\ref{sec:agent}.

The second stage of the workflow is simulation and visual testing. For this purpose, the VGA Playground UI \cite{b13} is used. The application combines RTL simulation based on the open-source tool \textit{Verilator} with a VGA rendering engine to produce interactive visual output. Users paste the RTL code into the tool, which then simulates the design in real time. The immediate visual feedback allows users to validate correctness and debug their designs without writing complex testbenches. Users can return to the chat agent to describe discrepancies between intended and observed behavior, iteratively refine their prompts, and regenerate the RTL code. 

The last stage of the workflow is the chip implementation. This step is based on the GitHub web UI and the Tiny Tapeout Verilog template \cite{b9}. Users insert their RTL code into the designated files, and upon pushing the changes, the logic synthesis and the backend flow are initiated based on GitHub Actions. Using the open-source tools \textit{Yosys} and \textit{LibreLane}, this step ultimately produces a GDSII file. The resulting layout can be inspected with the Tiny Tapeout GDS viewer \cite{b14}.

\subsection{LLM-based Chat Agent System}\label{sec:agent}

\begin{figure}[htbp]
\centerline{\includegraphics[width=0.5\textwidth]{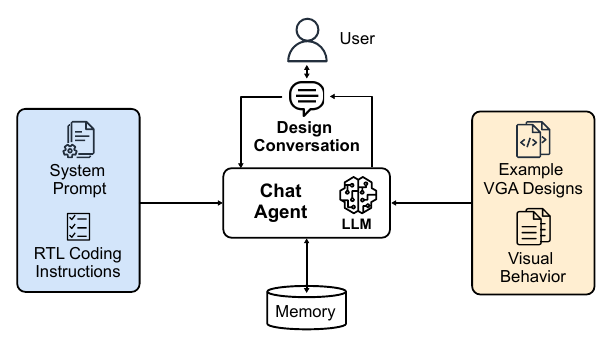}}
\caption{Overview of the LLM-based chat agent system.}
\label{fig:agent}
\end{figure}

Figure~\ref{fig:agent} shows the detailed configuration of the agent system. Users interact with the chat agent in a conversational manner through the UI described in the previous section. The system employs a single-agent setup based on an LLM. This design choice reflects the focused nature of the task, which centers exclusively on Tiny Tapeout VGA designs. In contrast, multi-agent systems are typically advantageous when the search space is broad and requirements are ambiguous. The system memory maintains the conversation history, intermediate design results, and contextual information to ensure consistency across multiple design iterations. 

To guide the agent’s operation, a detailed system prompt and RTL coding instructions are provided. The system prompt defines the agent’s role and constraints. It introduces the context of RTL coding in Verilog for VGA chip designs, describes the educational orientation of the task, and specifies the novice experience level of the users. Furthermore, it directs the agent to explain concepts in simple language and to generate extensively commented RTL code.

The RTL coding instructions provide a prescriptive workflow to the agent that produces designs compliant with the VGA Playground and the Tiny Tapeout backend flow. The instructions provide step-by-step coding guidance and enforce fixed module and signal declarations that satisfy the interface requirements. The central part of Tiny Tapeout VGA designs is the instantiation of the open-source VGA timing controller intellectual property (IP) core. The instructions ensure that the necessary wires are declared, and signal routing as well as output pin mapping are completed. The agent is instructed to extend this default module template with the user-specified behavior, inserting the required control and datapath logic.

In addition, domain-specific knowledge is integrated into the agent system through pairs of example VGA designs and textual descriptions of their corresponding visual behavior. These pairs of input (description) and expected output (RTL code) serve as references for the agent system. The examples are default designs provided in the VGA Playground.

\section{Case Study: High-School Student Evaluation}
The goal of the case study is to assess the feasibility of the proposed LLM-assisted design methodology in a real educational setting. We aim to evaluate if non-experts can complete the full design flow within a limited time frame. The primary evaluation metrics are the success rate in terms of (1) functional correctness, i.e., passing behavioral simulation and (2) tapeout readiness, i.e., the successful GDSII generation with the Tiny Tapeout backend flow. Furthermore, the case study should provide insights into the achievable design scale and complexity under the time and experience constraints. 

\begin{table*}[t]
\centering
\caption{Overview of student designs created during the case study.}
\label{tab:designs}
\begin{tabular}{c l c c c c c c c}
\hline
\textbf{Group} & \textbf{Design Description} & \textbf{Category} & \textbf{SLOC} & \textbf{Tiles} & \textbf{Creativity} & \textbf{Complexity} & \textbf{Functionally Correct?} & \textbf{Tapeout-ready?} \\
\hline
1 & Blue car driving on the street & Interactive & 84 & 1$\times$1 & Medium & Medium & Yes & Yes \\
2 & Aquarium with swimming fish & Animation & 89 & 1$\times$1 & High & Medium & Yes & Yes \\
3 & Pixel-art cat figure & Static sprite & 66 & 1$\times$1 & Medium & Low & Yes & Yes \\
4 & Blue square & Static sprite & 41 & 1$\times$1 & Low & Low & Yes & Yes \\
5 & Jumping stick figure & Interactive & 76 & 2$\times$2 & High & Medium & Yes & Yes \\
6 & Red car driving on the street & Interactive & 59 & 1$\times$1 & Medium & Medium & Yes & Yes \\
7 & Unicorn catching a carrot & Animation & 126 & 1$\times$1 & High & High & Yes & Yes \\
8 & Tree with falling leaves & Animation & 97 & 1$\times$2 & High & High & Yes & Yes \\
\hline
\end{tabular}
\end{table*}

\subsection{Experimental Setup}
The case study was conducted with 18 students from German high schools, spanning grades 10 to 12. The participants were organized into eight groups of two to three students. The total duration of the study was 90 minutes. The first 15 minutes were dedicated to an introduction covering the chip manufacturing process and the chip design flow. The remaining 75 minutes were allocated to hands-on design work using the pre-configured workflow shown in Fig.~\ref{fig:workflow}.

\subsection{Results and Observations}

\begin{figure}[htbp]
\centerline{\includegraphics[width=0.45\textwidth]{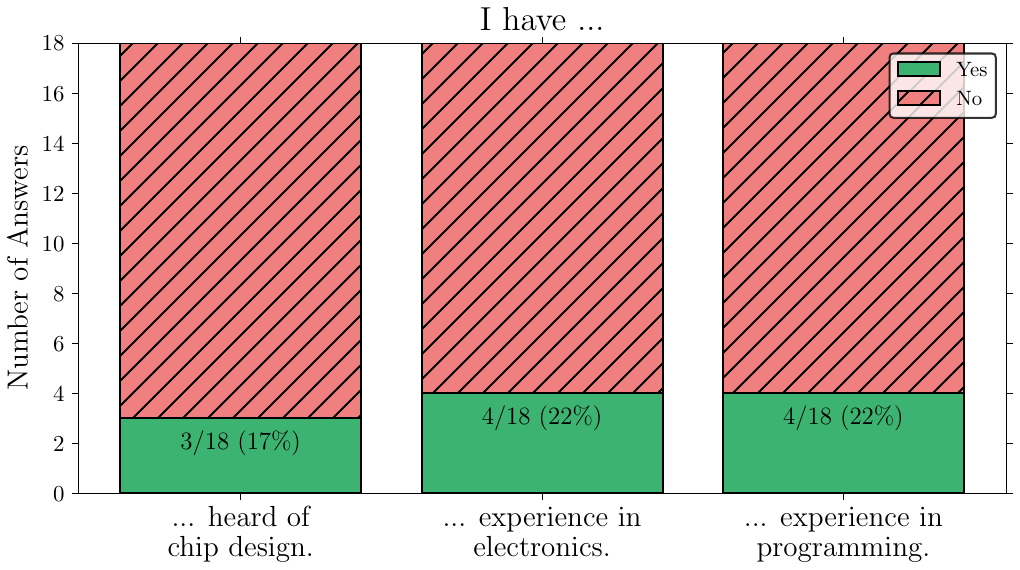}}
\caption{Results of the survey on students' background knowledge.}
\label{fig:survey}
\end{figure}

Before starting the design phase, a survey was conducted to assess the students’ prior experience. The survey included whether the participants knew what chip design is, and whether they had previous exposure to electronics (e.g., discrete components, soldering, do-it-yourself projects) or to programming. The responses are summarized in Fig.~\ref{fig:survey}. 

Three students (17\%) had heard of chip design, while four students (22\%) had first experience with electronics and the same number (22\%) with programming. The reported programming languages were Python, Java and HTML. The results indicate that the majority of students entered the case study without significant prior exposure to hardware or coding.

Table~\ref{tab:designs} summarizes the outcomes of the student projects. The designs are grouped into three categories: static sprite displays, non-interactive animations, and interactive elements controlled through the input pins. To provide an estimate of project scale, we report the effective source lines of code (SLOC) for each design, excluding comments and blank lines. In addition, the required silicon area is indicated in terms of the number of tiles, the smallest unit of integration in the Tiny Tapeout multi-project dies. Each tile measures approximately $160 \times 100\,\mu$m. Creativity and complexity were evaluated by two external experts based on the design idea and the RTL implementation. Finally, the table also notes whether VGA simulation and chip implementation were successful.

The eight student groups produced a diverse set of designs. Importantly, all designs successfully passed VGA simulation and the Tiny Tapeout backend flow.  Two groups implemented static sprite displays, which represent the simplest form of output and required between 41 and 66 SLOC. Three groups developed animations without user interaction, with source code sizes ranging from 89 to 126 lines. The remaining three projects implemented interactive elements, such as controllable cars or a jumping stick figure, with code sizes between 59 and 84 lines. In terms of hardware usage, most designs fit within a single tile. The more elaborate interactive stick figure required four tiles in total, and the falling tree animation extended to two tiles. Creativity and complexity varied considerably. While static sprites were ranked low to medium, animations and interactive applications reached medium to high ratings. These results demonstrate the feasibility of completing realistic and tapeout-ready projects within a short timeframe.

Beyond the quantitative results summarized in Table~\ref{tab:designs}, several qualitative insights emerged from the case study. Students were able to follow the generated RTL code by relying on the inserted comments. Those with prior programming experience used this approach to verify that the generated code aligned with their intended design behavior. Basic debugging strategies emerged naturally during the process, and the iterative, conversational design flow proved to be an effective way for students to incrementally build and refine their projects. For instance, the car projects evolved step by step from a simple square to a controllable vehicle, while the cat figure was gradually refined by successively adding visual features.

Challenges arose for the more complex designs. The tree animation consistently passed simulation but contained non-synthesizable Verilog code, requiring several cycles of backend error reporting and corrections using the chat agent. Similarly, both the stick figure and tree design exceeded the area of a single tile. This error could be solved by adjustments to the Tiny Tapeout configuration file based on the backend flow reports. However, the chat agent alone was insufficient for this task and instructor expertise was essential. These experiences highlight both the effectiveness of the conversational methodology and the current limitations in handling advanced backend issues.

\section{Conclusion and Future Work}
In this paper, we explored how LLMs can be leveraged to make chip design more accessible to beginners. We introduced an LLM-based chat agent integrated into a browser-based learning workflow built upon the Tiny Tapeout ecosystem. This approach enables users to progress from an initial design idea through RTL code generation to a tapeout-ready chip within a short time frame. A case study with 18 high-school students demonstrated that participants without prior experience were able to implement functional designs. All eight student groups successfully developed VGA chips in a 130\,nm technology during a 90-minute session. The resulting designs show the creative potential and technical feasibility of the methodology. While Tiny Tapeout has already enabled over 2,000 designs within three years, our approach further broadens this reach by empowering an even larger community of non-experts to easily enter the field of chip design.
Future work will focus on the extension of the chat agent and workflow to other audiences and the further analysis to quantify the users' learning curves.

\end{document}